\documentstyle{l-aa}

\begin{document}

\thesaurus {11.07.01, 11.09.2, 11.16.1, 11.16.2  }

\title{Candidate polar-ring galaxies in the Hubble Deep Field}

\author{V.P.~Reshetnikov\inst{1,2}}

\offprints{V.~Reshetnikov, resh@aispbu.spb.su}

\institute{ Astronomical Institute of St.Petersburg State University,
            198904 St.Petersburg, Russia
\and
            DEMIRM, Observatoire de Paris, 61 Av. de l'Observatoire,
            F-75014 Paris, France}

\date{Received 1996; accepted}

\maketitle

\begin{abstract}
We discuss the properties of two peculiar galaxies (2-809 and 2-906) 
selected in the Hubble Deep Field as possible candidates to
high-redshift ($z\approx~$1) polar-ring galaxies. 
We found that the presence of polar-ring galaxies in a random
deep field gives some support for a galaxy interaction rate 
steeply increasing with redshift.


\end{abstract}

\section{Introduction}

Polar-ring galaxies (PRGs) are among the most unusual and rare
extragalactic objects. PRGs are early-type galaxies with a 
large-scale ring of gas, dust and stars orbiting nearly perpendicular
to the major axis of the host galaxy. According to Whitmore
et al. (1990) (PRC) about 0.5\% of all nearby S0 galaxies are
observed with a polar ring. After correction for various selection
effects, the fraction increases to about 5\% of S0 galaxies
which have, or have had a polar ring.

An existence of two almost orthogonal kinematic systems strongly
suggests that PRGs are formed in galactic interactions.
The two most straightforward mechanisms of the polar-ring formation -
the accretion of matter from a nearby gas rich galaxy and 
the capture and merging of a companion - were recently 
illustrated by means of numerical simulations (Weil \& Hernquist
1993, Sotnikova 1996, Reshetnikov \& Sotnikova 1997). 
Besides, there are several binary systems
in the local universe clearly demonstrating polar-rings formation
in galactic interactions (e.g., Reshetnikov et al. 1996).

It is usually believed that galaxy interactions were more frequent 
in the past (e.g., Keel 1996 for a recent review).
Therefore, one can expect to find a larger fraction of PRGs   
among high-redshift objects (assuming that the lifetime of
polar rings is limited). 
Comparison of the PRGs fraction in the local universe with 
that among high-redshift galaxies will provide an estimate
of the rate of this specific type of interactions (leading to
polar-ring formation) as a function of redshift. 
Note that recent statistics of distant $ring$ galaxies provided
strong evidence for a very steep increase of a galaxy 
interaction rate with redshift (Lavery et al. 1996).   

The Hubble Deep Field (HDF) (Williams et al. 1996) images,
obtained with the Hubble Space Telescope during
December 1995, permit to study distant
galaxies in unprecedented details. 
In this note we present two polar-ring candidates found in 
the HDF and discuss possible consequences of the PRGs presence
in the HDF for the evolution of interaction rate.

\section{PRGs in the Hubble Deep Field}

After visual inspection of high resolution (3069 $\times$ 3100 pixels)
colour images of the HDF stored on the ST ScI world-wide web site,
we selected two
relatively bright galaxies - 2-809 and 2-906 - as the most promising
candidates in high-redshift PRGs. 
The galaxies do not have measured spectroscopic redshifts.

The global photometric characteristics of the galaxies according
to Williams et al. (1996) are summarized in
the Table 1. The columns of the table are: galaxy identification; 
coordinates (epoch J2000); photometric redshift (Lanzetta et al. 1996);
``total'' magnitude\footnote{All magnitudes in the paper are in the
$AB$ system (Oke 1974): $m_{AB}=\rm -2.5$log$f_{\nu}-$48.60,
where $f_{\nu}$ is the flux density 
in units of $\rm erg~s^{-1}~cm^{-2}~Hz^{-1}$} 
in the F606W filter ($v$ magnitude);
isophotal colours (F300W$-$F450W as $u-b$, F450W$-$F606W as $b-v$, and
F606W$-$F814W as $v-i$)
measured to a faint limiting isophote determined
from the summed F606W+F814W image.

\begin{table*}
\caption[1]{General characteristics of the galaxies}
\begin{tabular}{llllllll}
\\
\hline \\
ID & $\alpha$(m:s) & $\delta$(\arcmin:\arcsec) & z &  $v_{t}$ & $u-b$&
$b-v$ & $v-i$ \\ 
\\
\hline \\
2-809 & 36:52.88 & 14:05.1 & 0.64 & 24.05  & 0.53 & 0.78 &
0.79 \\
2-906 & 36:56.13 & 13:29.7 &      & 24.00  & 0.54 & 0.23 &
0.56 \\
\\
\hline
\end{tabular}
\\
\end{table*}

In order to study polar-ring candidates,
we retrieved and processed the HDF images (produced by the Version 2
drizzle algorithm) of the galaxies in the MIDAS environment.

\subsection{2-809}

In the Fig.1 we present a reproduction of the galaxy from the combined
F606W+F814W frame (left side of the figure). Right side of the
figure shows DSS\footnote{The Digitized Sky Surveys were produced at
the Space Telescope Science Institute under U.S. Government grant
NAG W-2166.} image of the optical counterpart of 2-809 among
local PRGs (NGC 4650A). As one can see, 2-809 resembles remarkably
the well-known polar-ring galaxy NGC 4650A.  

\begin{figure}
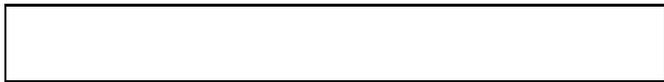

\picplace{1cm}
\caption[1]{Deep (F606W+F814W) image of 2-809 (left). The size of the
image is 3\arcsec$\times$6.6\arcsec. Right $-$ Digital Sky Survey image
of NGC 4650A (2\arcmin$\times$4.7\arcmin.) Both images are in arbitrary
orientation. }
\end{figure}

Fig. 2 shows isophotal contour map of 2-809 constructed from
the summed F606W+F814W frame. 
Two warped extended structures are stretched along the minor
axis of the galaxy out to at least $r=$2\arcsec. 
The morphological similarity between the galaxy and the
classic PRGs (see, for instance, PRC) is evident in this
figure. The east side of the suspected edge-on ring is
extended towards peculiar galaxy 2-856. The projected
distance between 2-809 and 2-856 is 4.4\arcsec or 26 (23) kpc
for $H_{0}=75$ km/c/Mpc, $z=$0.64 and $q_{0}=$0.05 (0.5). There is
a small extention from NW side of 2-856 directed
to 2-809. This feature may be interpreted as a consequence of
tidal interaction between 2-856 and 2-809. Therefore, 2-856 may be 
considered as a possible ``donor'' galaxy - the stripped matter 
from this galaxy could form the ring around 2-809.
  
\begin{figure}
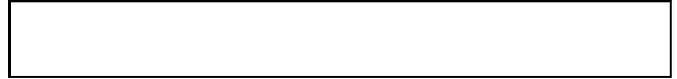

\picplace{1cm}
\caption[2]{Isophotal contour map of 2-809 constructed from the 
summed F606W+F814W frame. The faintest contour corresponds to
2$\sigma$ of the sky level, isophotes step is $\rm 0.^{m}75$.
The arrow indicates north, east is $\rm 90^{o}$ counterclockwise.
The arrow length is 2\arcsec.}
\end{figure}

Fig.3 presents the $v$ and $i$ photometric cuts along the major
axis ($\rm P.A.=174^{o}$) of the main body of 2-809 and 
along the bright inner part of the suspected ring ($\rm P.A.=81^{o}$).
The main body demonstrates peculiar surface brightness
distribution. In the $i$ passband the galaxy shows
almost constant surface brightness within 0.3$\arcsec$ from
the nucleus with the following fast decreasing of the brightness. 
In the $v$ filter there is a minimum in the center of
the galaxy resulted in local reddening in this region.
Then, the $v-i$ colour systematically increases to the 
galactic periphery. On the contrary, the $b-v$ colour shows sharp
minimum in the center (up to $\approx$+0.6) with the following
increasing to 1-1.5. One can propose that we observe here
the projection of the blue ring on the central 
region of the galaxy. The suspected
ring is notably bluer in comparison with central galaxy (Fig.3b).
     
\begin{figure}
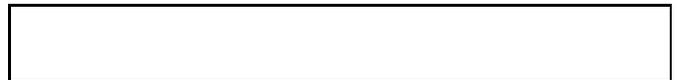

\picplace{1cm}
\caption[3]{Liminosity profiles (in $\rm mag~arcsec^{-2}$) along
the major axis of the central galaxy of 2-809 (a) and along the
suspected polar ring (b) in the $i$ (solid line) and $v$ (dashed line)
passbands.}
\end{figure}

The main observational characteristics of the galaxy are 
summarized in the Table 2 for $q_{0}=$0.05 (0.5) and
$H_{0}=$75 km/s/Mpc. The total magnitudes and colours in
the table were determined from the multi-aperture photometry.
The results of our photometry are in good agreement with
that of Williams et al. (1996). The observed characterstics
of the central galaxy in the table are corrected for the ring
contribution. The central galaxy demonstrates
relatively red colours while the suspected ring is significantly
bluer in the $b-v$ colour. The difference in observed
colours between the central galaxy and the ring is typical for
local PRGs (e.g., Reshetnikov et al. 1994).  

Approximate absolute ``blue'' magnitudes were found from the apparent
$b$ magnitudes using the $K$-correction from Cowie et al. (1994)
for a ``Sa-Sb'' galaxy at this redshift for the main galaxy
and for a ``Sbc-Sdm'' galaxy for the suspected ring. 
The correction  
between the instrumental $b$ band and the standard $B$ filter 
is used also according to Holtzman et al. (1995).
Absolute luminosities of the central galaxy and the ring of
2-809 are close to that of local PRGs (Schweizer et al. 1983,
Whitmore et al. 1987, Reshetnikov et al. 1994).
(It should be noted that
the derived absolute magnitudes must be considered as tentative
only due to uncertain redshift and $K$-correction for the
galaxy.)

Summarizing our analysis of the 2-809 photometric characteristics,
one can conclude that in general this galaxy resembles well
local PRGs with extended polar rings (Schweizer et al. 1983,
Whitmore et al. 1987, Reshetnikov \& Sotnikova 1997). 
As an alternative explanation of the 2-809 optical morphology
one can propose that this galaxy is a ``NGC 7252-like'' merger with 
two symmetric tidal tails. Future spectral observations must
clear up the true nature of this peculiar galaxy. 

\begin{table}
\caption[2]{Derived characteristics of the polar-ring candidates}
\begin{tabular}{lll}
\\
\hline \\
 & 2-809 & 2-906 \\ 
\\
\hline \\
Adopted redshift        & 0.64          & 1.2            \\
{\it Central galaxy} \\ 
Total magnitude ($i$)   & 23.6          & 23.85          \\
Absolute blue magnitude & -18.6: (-18.3:)& -21.5: (-21.0:)\\
$v-i$                   & +0.9          & +0.6           \\
$b-v$                   & +1.1          & +0.25          \\
Diameter ($\mu_{i}=$26) & 9.6 (8.3) kpc & 11.5 (8.9) kpc \\
Position angle          &  $\rm 174^{o}$: & $\rm 124^{o}$ \\
Axial ratio             & 0.6:          & 0.4:               \\
\\
{\it Suspected ring} \\
Total magnitude ($i$)   & 24.4          & 24.6           \\
Absolute blue magnitude & -17.4: (-17.1:)& -19.3: (-18.7:)\\
$v-i$                   & +0.8          & +0.6           \\
$b-v$                   & +0.4          & +0.1           \\
Diameter ($\mu_{i}=$26) & 31: (27:) kpc & 11.5 (8.9) kpc \\
Position angle          & $\rm 81^{o}$: & $\rm 48^{o}$   \\
Ring to main galaxy\\
luminosity ratio ($i$)  &  0.5          &  0.5           \\
\\
\hline
\end{tabular}
\end{table} 

\subsection{2-906}

Fig.4 compares the morphology of the HDF galaxy 2-906 with the
prototype polar-ring galaxy NGC 2685. The similarity between
the galaxies is noticeable. The suspected ring in 2-906
has approximately the same diameter as the main galaxy and
is inclined by about $\rm 75^{o}$ to its plane. The apparent 
axial ratio of the ring structure is about 0.5. 

\begin{figure}
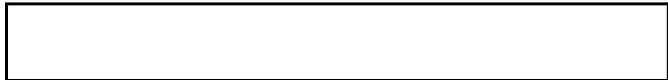

\picplace{1cm}
\caption[4]{Deep (F606W+F814W) image of 2-906 (left). The size of the
image is 3.2\arcsec$\times$3.5\arcsec. Right $-$ Digital Sky Survey image
of NGC 2685 (2.3\arcmin$\times$2.5\arcmin.) Both images are in arbitrary
orientation. }
\end{figure}

The galaxy has no published spectroscopic or photometric
redshift (it is located near the edge of the chip and was 
omitted by Lanzetta et al. 1996). In order to estimate the redshift,
we used a ``differential'' method: we chose galaxies with
determined photometric redshifts from Lanzetta et al. (1996)
in the interval $\rm 23.0\leq${\it i}$\leq24.0$; selected in this
sample the galaxies with ``exponential-like'' surface brightness 
distribution in the $i$ passband similar to surface brightness
distribution in 2-906 (see further); we considered for these
galaxies relationships between observational colours and
estimated redshift and examined the 2-906 characteristics
position. We found distinct non-linear
dependence between $z$ and colour $b-v$ for the sample galaxies
(see Fig.5) and from the integral colour of the galaxy
we estimate the redshift as 1.2. (This redshift does not
contradict the other colours dependences on $z$ also.)
Naturally, this redshift estimation is very uncertain and
we will use it with illustrative purpose only.  

\begin{figure}
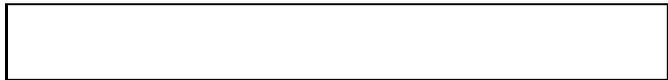

\picplace{1cm}
\caption[5]{Dependence of the observed colour $b-v$ for the
HDF galaxies with exponential-like surface brightness distribution
and total magnitude $\rm 23.0\leq${\it i}$\leq24.0$ on the photometric
redshift according to Lanzetta et al. (1996) (circles).
The solid line segment shows location of the 2-906
colour, the star represents 2-809.}
\end{figure}

Fig.6 presents isophotal contour map of the galaxy. 
The central galaxy shows peculiar structure -
asymmetric surface brightness distribution and warped
outer regions. In the $i$ passband the galaxy reveals
exponential-like surface brightness 
distribution (see Fig.7) with exponential scalelenght of about 
0.33\arcsec or 2.4 (1.9) kpc. This is typical value for present-day
spiral galaxies (e.g., de Jong 1996). In the $v$ passband
the surface brightness is almost flat in the central
region, which results in a strong $v-i$ colour
gradient from the nucleus to galaxy periphery. The photometric
cut along the major axis of the suspected ring (Fig.7b)
shows the  asymmetry of the ring and its bluer
colour in comparison with the central galaxy. 

\begin{figure}
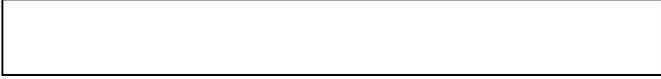

\picplace{1cm}
\caption[6]{Isophotal contour map of 2-906 from the combined
F606W+F814W frame. The faintest contour corresponds to 2$\sigma$
of the sky level; the fist three isophotes are separated by
$\rm 0.^{m}5$, then the isophotes step is $\rm 0.^{m}3$. The
arrow indicates north, east is $\rm 90^{o}$ counterclockwise.
The arrow length is 1\arcsec.} 
\end{figure}

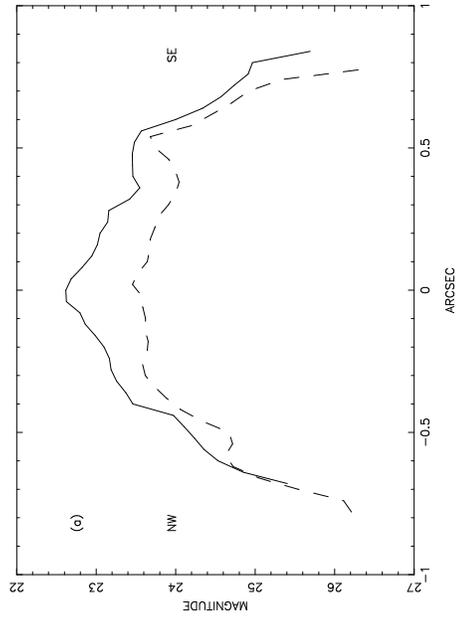
\begin{figure}
\picplace{1cm}
\caption[7]{Luminosity profiles (in $\rm mag~arcsec^{-2}$) along
the major axis of the central galaxy of 2-906 ($\rm P.A.=124^{o}$)
(a) and along
the suspected polar ring ($\rm P.A.=48^{o}$)
(b) in the $i$ (solid line) and $v$ (dashed 
line) passbands.} 
\end{figure}

The general observational characteristics of the galaxy are summarized
in Table 2. The main properties of 2-906 are quite
consistent with the characteristics of local PRGs with relatively
short polar rings (Schweizer et al. 1983, Whitmore et al. 1987,
Reshetnikov \& Sotnikova 1997). One can 
mention only a somewhat brighter absolute luminosities of the
galaxy and the ring (absolute blue luminosities were estimated 
by the same manner as for 2-809 - see section 2.1). 
This can be explained by uncertaintes in adopted redshift, $K$-correction
and, probably, by active star formation in the central galaxy
of 2-906.

\section{Discussion}

As it was shown in the previous section, the two HDF polar-ring candidates
resemble well the general photometric and morphological properties of 
local PRGs. The main difference between local and distant PRGs is the 
structure of the host galaxy. 
Both polar-ring candidates show peculiar main 
bodies while local PRGs are usually gas-free early-type galaxies. 
One can propose that the observed central galaxies peculiarities 
are due to strong tidal shaking and matter accretion during
the rings formation. A few billions years later - in the present
epoch - after the gas consumption in star formation, the central
galaxies become early-type looking galaxies. 

The presence of two good polar-ring candidates in the small HDF
is very interesting and probably provides some information about
the frequency of galactic interactions at middle ($z\approx$~1) redshifts.
Folowing Lavery et al. (1996), we estimate the likelihood of finding
a single polar-ring galaxy in the HDF for two models: (1) a polar-ring
galaxy volume density does not change with 
redshift (the nonevolving
case) and (2) a polar-ring volume density increases with redshift
(the evolving case). 

In our calculations we have assumed that PRGs lie between the redshifts
0.5 and 1.5. According to PRC, 0.5\% of all
local S0 galaxies appear to have polar rings. (We did not correct
this estimation for selection effects since in the 
first-order approximation
the same selection effects must reduce the observed fraction of PRGs
in the HDF also. Note, for instance, that the HDF galaxies are 
observed with the same orientation as local PRGs - edge-on central
galaxies and almost edge-on suspected rings. See discussion about
polar ring recognizion in the PRC.) Therefore, adopting that local
S0 galaxies constitute about 1/10 of all galaxies and that the volume
density of non-dwarf galaxies in the present epoch is 
0.01-0.02 $\rm Mpc^{-3}$,
we obtain the estimation of the local PRGs volume density as
$\rm (5-10)\times10^{-6}$ $\rm Mpc^{-3}$. 

Then, varying $H_{0}$ from 50 to 100 km/s/Mpc and $q_{0}$ from 0.05
to 0.5, we found that the probability of detecting a single PRG
in the HDF is, on average, $\rm 0.06\pm0.02$ only for the nonevolving 
case. In the second 
case, assuming that the polar-ring galaxy volume density increases
with redshift as $\rm (1+$z$)^{4-4.5}$, we found that the expected
number of PRGs is $\rm 1.6\pm1.0$.
Thus, in the evolving case 
the expected number of galaxies is consistent with observations.
Therefore, the PRGs presence in the HDF suggests
that the volume density of PRGs does increase with redshift
and provides evidence of a steeply increasing galaxy
interaction rate (leading to a polar ring formation) up to $z\approx$1. 
This conclusion is in accordance with recent statistics of
distant ring galaxies. According to Lavery et al. (1996),
a surprisingly large number of ring galaxy candidates
in a random field suggests a rapid increase of
a galactic interaction rate with redshift. (Let us note also
that our adopted local volume density of PRGs is close to that
for ring galaxies according to Few \& Madore 1986. Therefore,
one can expect to find a several ring galaxies in the HDF also.)

Another way (more rough but straightforward) of estimating
the fraction of PRGs in the HDF - ratio the number of candidates
by the number of galaxies in the HDF that are resolved enough to
identify a polar ring - leads to the same conclusion about
evolution of the volume density of PRGs. The fact that both
candidates are 24th magnitude ($v$ filter) suggests that we
may not be able to identify PRGs much fainter than that.
Assuming that we can identify a polar ring in all HDF galaxies
with $v~<~$25.5, we have relative fraction of PRGs as 
$\approx$2/300~$=$~0.7\%. This relative fraction is significantly
higher than local value of about 0.05\% (Whitmore et al. 1990).

It should be noted, however, that present statistics of PRGs
must be considered with caution. Both HDF candidates
must be confirmed as true PRGs by kinematical observations.  
Moreover, the HDF polar rings are, probably, relatively
young objects (from asymmetry of surface brightness and
colour distributions) while among local PRGs there are
old (for instance, A 0136-0801) and more recently formed rings
(probably, NGC 4650A). Some local PRGs may be formed several
$\rm 10^{9}$ yr ago and, therefore, the statistics of present-day
polar rings do not reflect local rate of galaxy interactions.
Taking into account relatively young local PRGs only, we will
have an evidence of even more rapid increase of the galactic
interaction rate with $z$. Some additional limitations 
of such approach - estimation of the galactic interaction
rate at high $z$ from the statistics of a specific interaction
relics - are discussed by Lavery et al. (1996).

\section{Conclusions}

Visual inspection of the HDF has revealed two relatively
bright peculiar objects (2-809 and 2-906) morphologically 
resembling local polar-ring galaxies. We find that these
objects are very similar to classic PRGs by general
photometric characteristics. The presence of these
objects in the HDF suggests that the volume density of
PRGs does rise with redshift and supports a steeply increasing
galaxy interaction rate at $z\approx$1.

\acknowledgements{I would like to thank Fran\c coise Combes and
anonymous referee for useful comments.
I acknowledge support from French Ministere 
de la Recherche et de la Technologie during my stay in Paris.
This work was supported by grants $N$ 94-2-06026-a and
95-02-05596 from Russian Foundation for Basic Research.}

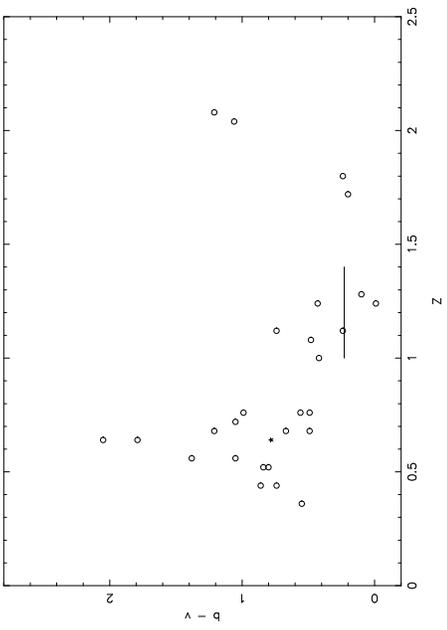
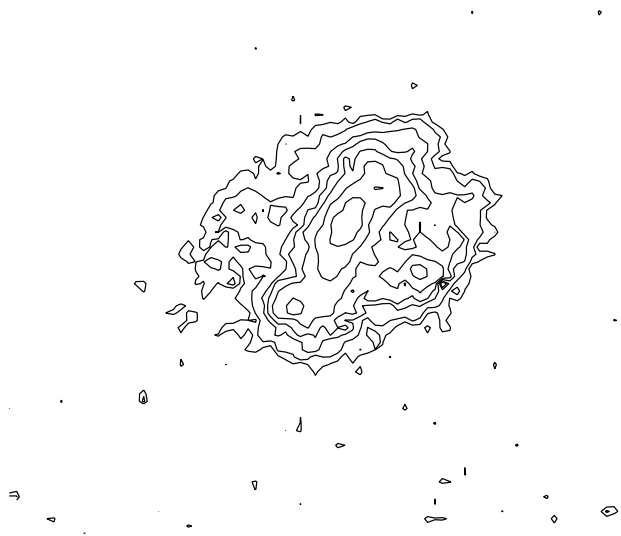
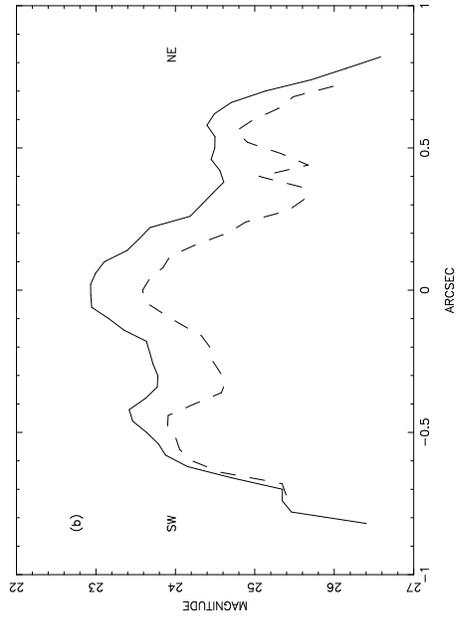
\end{document}